\pdfoutput=1

\documentclass[reprint,aps,prl,twocolumn,,floatfix,superscriptaddress]{revtex4-1}
\usepackage{graphicx}
\usepackage{amssymb}
\usepackage{dcolumn}
\usepackage{bm}
\usepackage{amsmath}
\usepackage{lineno}
\usepackage[english]{babel}
\usepackage[utf8]{inputenc}
\usepackage{siunitx}
\usepackage{xcolor}
\RequirePackage[T1]{fontenc}
\RequirePackage{lmodern}

\begin{document}

\title{Valley splitting in silicon from the interference pattern of quantum oscillations}

\author{M. Lodari}
\affiliation{QuTech and Kavli Institute of Nanoscience, Delft University of Technology, PO Box 5046, 2600 GA Delft, The Netherlands}
\author{L. Lampert}
\affiliation{Intel Components Research, Intel Corporation, 2501 NW 229$^{th}$ Avenue, Hillsboro, OR, USA}
\author{O. Zietz}
\affiliation{Intel Components Research, Intel Corporation, 2501 NW 229$^{th}$ Avenue, Hillsboro, OR, USA}
\author{R. Pillarisetty}
\affiliation{Intel Components Research, Intel Corporation, 2501 NW 229$^{th}$ Avenue, Hillsboro, OR, USA}
\author{J.S. Clarke}
\affiliation{Intel Components Research, Intel Corporation, 2501 NW 229$^{th}$ Avenue, Hillsboro, OR, USA}
\author{G. Scappucci}
\email{g.scappucci@tudelft.nl}
\affiliation{QuTech and Kavli Institute of Nanoscience, Delft University of Technology, PO Box 5046, 2600 GA Delft, The Netherlands}

\date{\today}
\begin{abstract}
We determine the energy splitting of the conduction-band valleys in two-dimensional (2D) electrons confined in silicon metal oxide semiconductor (Si-MOS) Hall-bar transistors. These Si-MOS Hall bars are made by advanced semiconductor manufacturing on 300~mm silicon wafers and support a 2D electron gas of high quality with a maximum mobility of $17.6 \times 10^3$~cm$^{2}$/Vs and minimum percolation density of $3.45 \times 10^{10}$~cm$^{-2}$. Because of the low disorder, we observe beatings in the Shubnikov-de Haas oscillations that arise from the energy-split two low-lying conduction band valleys. From the analysis of the oscillations beating patterns up to $T=1.7$~K, we estimate a maximum valley splitting of $\Delta E_{VS} = 8.2$~meV at a density of $6.8 \times 10^{12}$~cm$^{-2}$. Furthermore, the valley splitting increases with density at a rate consistent with theoretical predictions for a near-ideal semiconductor/oxide interface.
\end{abstract}

\maketitle
\begin{figure}[!ht]
\includegraphics{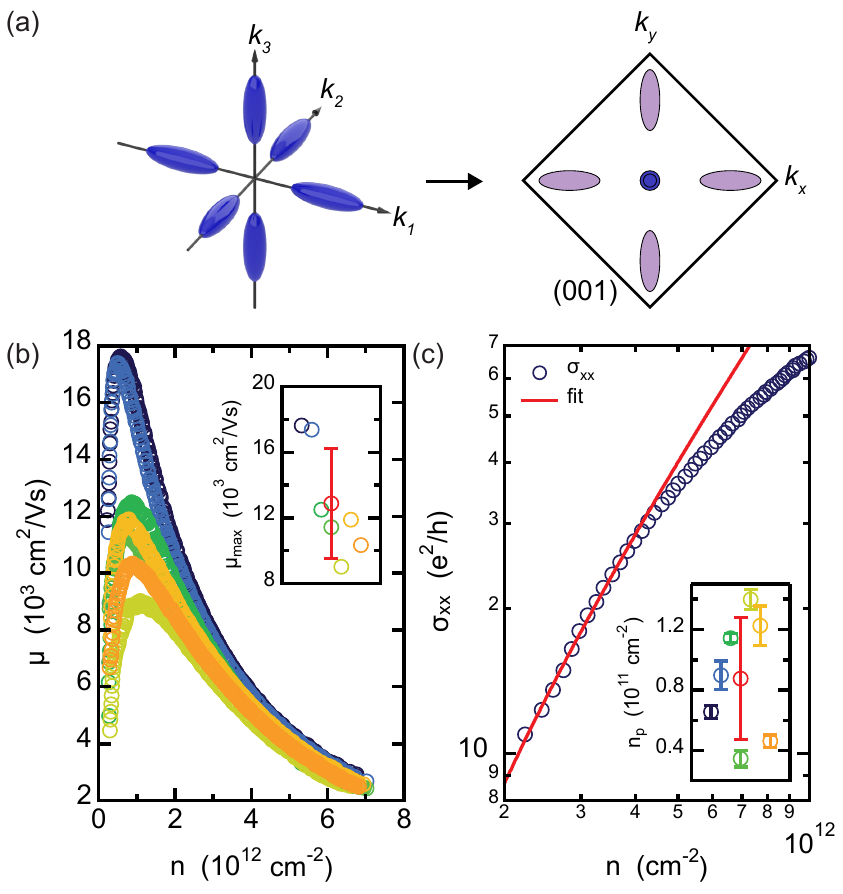}
\caption{\label{fig:1}(a) Schematic of the constant-energy ellipsoids for the Si conduction-bands in momentum space (left cartoon) and constant energy ellipses obtained by 2D projection for the (001) Si surface (right cartoon).  Lower energy subbands are shown in blue. Long and short axis of the  ellipsoids correspond to the longitudinal ($m_l=0.92m_0$) and transverse ($m_t=0.19m_0$) effective mass for electrons in Si, respectively. States from the out-of-plane valleys have the heaviest effective mass ($m_l$) along the quantization axis and form  a double-degenerate ground state in 2D (concentric ellipses), further split in energy by the sharp confinement potential. (b) Mobility $\mu$ as a function of density $n_H$ from Hall bar devices across a wafer at $T = 1.7$~K. Circles colored in black, blue, green, light green, yellow, orange and dark orange correspond to measurement from samples A-G, respectively. The inset shows the maximum mobility $\mu_{max}$ from all the samples and average value $\pm$ standard deviation (red). (c) Conductivity $\sigma_{xx}$ as a function of $n_H$ from device A (dark blue circles) with fit to percolation theory in the low density regime (red line). The inset shows the percolation density $n_p$ from all the devices and average value $\pm$ standard deviation (red).}
\end{figure}

Electron spin qubits in silicon quantum dots are a compelling candidate for quantum processors because they have long coherence time\cite{Zwanenburg_RevModPhys2013,Veldhorst_longCoherence_2014,Veldhorst_2qubit_2015,Lieven_semiconQC_2019,Yang_NatEl_2019,Huang_fidelity_2019}, can operate quantum logic above one Kelvin\cite{Yang2020_hotqubit,Petit_hotqubit_NAT}, thereby providing scope for integration of classical control electronics\cite{Xue_cryoCMOS_2021}, and leverage advanced semiconductor manufacturing\cite{zwerver2021qubits}. To accelerate the device fabrication/measurement cycle towards larger quantum processors, it is crucial to characterize with high throughput the key electrical properties of the material, such as mobility, percolation density, and valley splitting energy. Whilst mobility and percolation density are well established metrics to qualify disorder in materials hosting spin-qubits\cite{Sabbagh_industrial,PaqueletWuetz2020_cryomux}, measurements of valley splitting energy in silicon remains challenging. Quantum confinement across a (001)~interface removes the six-fold degeneracy of the conduction-band valleys in Si (Fig.~\ref{fig:1}(a))\cite{fang_negative_1966,stern_properties_1967}. A two-fold degenerate ground-state is formed from the two out-of-plane valleys that present the heavy longitudinal effective mass of Si oriented along the quantization axis. The remaining two-fold valley degeneracy is further lifted by the presence of a sharp confinement potential\cite{sham_effective-mass_1979,ando_electronic_1982} and valley splitting quantifies the relevant energy separation.

Valley splitting is measured in quantum dot (QD) devices or in Hall bars field effect transistors. In a typical QD measurement, the single-particle energy level splitting is obtained by monitoring the increase in spin relaxation at the hot-spot\cite{VS_Yang2013,Petit_hotqubits_PRL}. These measurements are critical for developing functional qubits and give important insights on the small-scale variation of valley splitting at the device level. Alternatively, measurements in Hall bars probe the energy splitting of the two low-lying conduction-band valleys in 2DEGs. Due to the different confinement experienced by electrons, Hall bar measurements are not a direct probe of valley splitting in qubits, but are still a useful quick turn monitor for high-throughput characterization and materials optimization. However, the evaluation of valley splitting in Hall bars relies on activation energy measurements in the quantum Hall effect (QHE) regime\cite{ando_electronic_1982}. Due to the large magnetic field needed to overcome the Landau level broadening, enhancement of energy gaps is observed\cite{TRACY_VS_enhancement_2010}, making a direct comparison to the single-particle energy levels of QDs challenging. Furthermore, the complex electrostatics of quantum Hall edge states must be taken into account to correctly interpret the measurements\cite{Paquelet_VS}.

In this Letter we determine valley splitting in silicon by analyzing the quantum interference properties of 2DEGs measured in magnetotransport. These measurements are performed in low-disorder Si-MOS Hall bars. At a density greater than $3.7 \times 10^{12}$~cm$^{-2}$, the large vertical electric field increases valley splitting above the disorder-induced single-particle energy level broadening, thereby making visible the subtle interference effects due to quantum transport through the two energy-split valleys.

The Si-MOS Hall bars are fully-optically patterned, feature a composite SiO$_2$/high-$\kappa$ thin dielectric and are made in a 300~mm wafer process line using the quantum dot process flow described in refs.~\cite{pillarisetty_si_2021,zwerver2021qubits}. The width of the Hall bars is 6~$\mu$m and the interval between voltage terminals is 30~$\mu$m.  Magnetotransport characterization was performed at $T=1.7$~K and at $T=65$~mK in refrigerators equipped with cryo-multiplexers \cite{PaqueletWuetz2020_cryomux} using standard four-probe low-frequency techniques with excitation source-drain bias of $1$~mV. A positive bias applied to the gate ($V_g$) induces a 2DEG at the semiconductor/oxide interface.

\begin{figure}[!ht]
\includegraphics{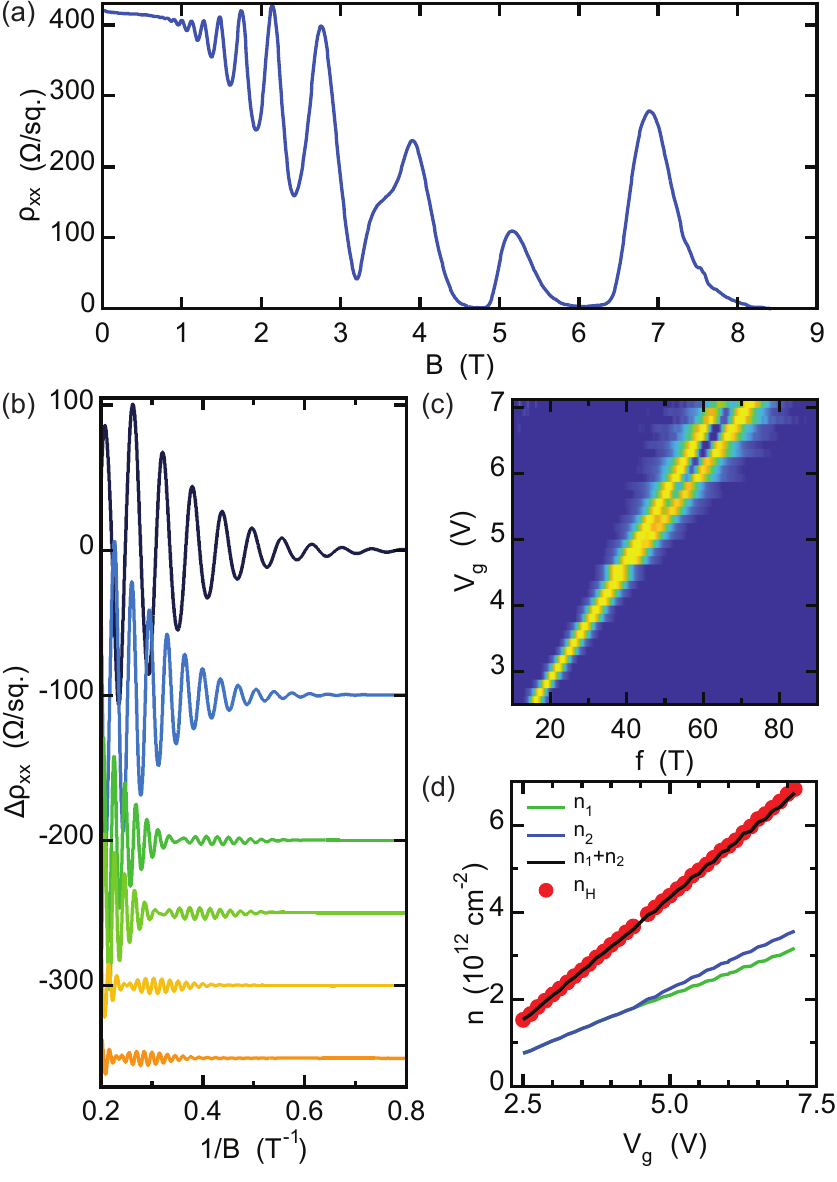}
\caption{\label{fig:2}(a) Longitudinal resistivity $\rho_{xx}$ for device B as a function of magnetic field $B$ at a Hall density $n_H = 9.3 \times 10^{11}$~cm$^{-2}$ and mobility $\mu = 16.1 \times 10^3$~cm$^{2}$/Vs at $T = 65$~mK. (b) Oscillation amplitude $\Delta\rho_{xx} = \rho_{xx} - \rho_{0}$, where $\rho_{0}$ is the low field resistivity, as a function of the inverse perpendicular magnetic field $1/B$ for device B after smoothing and polynomial background subtraction at $T = 1.7$~K. Different curves correspond to different and increasing accumulation gates $V_g$ (dark blue to orange, respectively): $V_g = 2.6$, $3.6$, $5,1$, $5.5$, $6.5$, and $6.9$~V corresponding to $n_H =1.7$, $2.8$, $4.5$, $5.0$, $6.1$, and $6.5\times 10^{12}$~cm$^{-2}$.
The curves are offset for clarity. (c) The normalized fast Fourier transform (FFT) spectra amplitude of the oscillations for device B as a function of accumulation gate $V_g$ and oscillation frequency $f$ at $T = 1.7$~K. Amplitude color scale : $0.3$ to $1$. To obtain the FFT we use the raw $\Delta\rho_{xx}$ data with no background subtraction. Smoothing and interpolation are performed by using a Savitzky-Golay Matlab smoothing routine to obtain a $1/B$ equally spaced signal to feed into the FFT. (d) Comparison of densities from Hall effect and FFT analysis of the SdH oscillations as a function of accumulation gate $V_g$. $n_1$ (green) and $n_2$ (blue) are the single valley densities from FFT and $n_{1}+n_{2}$ (black) is the resultant total density. $n_H$ is the Hall density (red circles).}
\end{figure}

\begin{figure}[ht]
\includegraphics{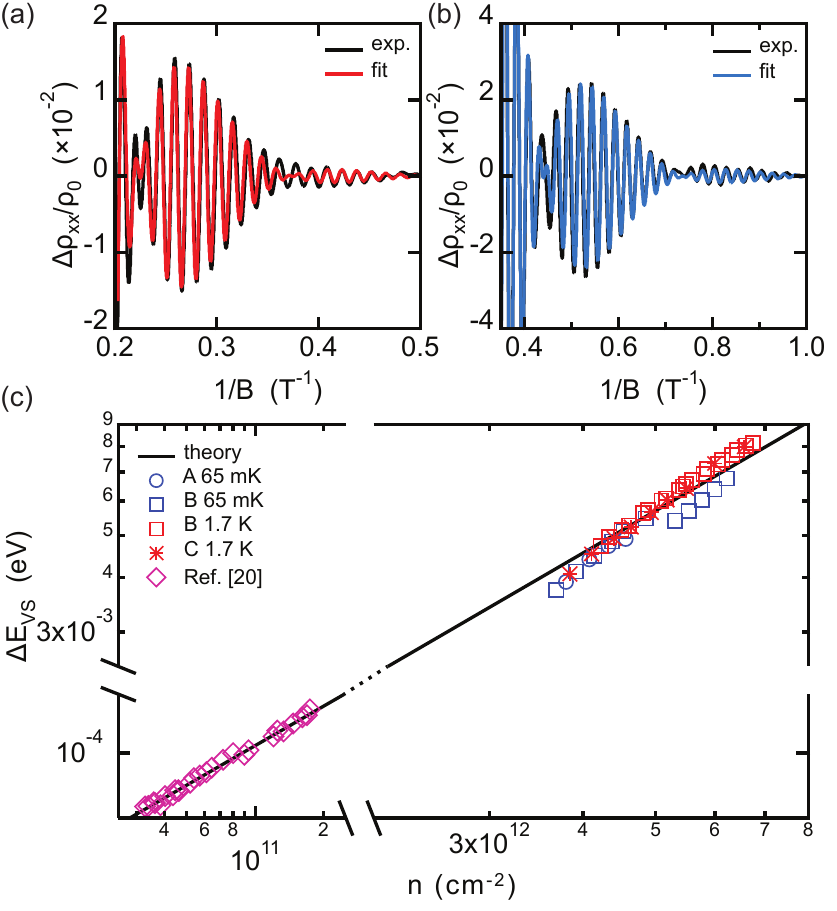}
\caption{\label{fig:3} Sample B oscillations amplitude normalized to the low-field magnetoresistance $\Delta\rho_{xx}/\rho_{xx,0}$ as a function of inverse perpendicular field $1/B$ after smoothing and polynomial background subtraction (black) and their fittings (red and blue) at a temperature of (a) $T = 1.7$ K and (b) $65$~mK. (c) Valley splitting $\Delta E_{VS}$ as a function of density $n$ at a temperature of $T = 1.7$~K (red) and $65$~mK (blue) from devices A (circles), B (squares) and C (asterisks). The black line is the theoretical density dependence of valley splitting $\Delta E_{VS} \sim 1.14 \times n$ where $\Delta E_{VS}$ and $n$ are in meV and $10^{12}$~cm$^{-2}$ units calculated in ref.~\cite{Fiesen2007VS_theory}.
Magenta diamonds are experimental valley splitting values from Si/SiGe heterostructure-FET from ref.~\cite{Paquelet_VS}}
\end{figure}

Figure~\ref{fig:1}(b) shows the mobility-density curves at $T = 1.7$~K for seven devices (A--G) across the same wafer. The mobility increases as a function of density, due to the increased screening of scattering from impurities, until a peak is observed ($\mu_{max}$). At higher density, surface roughness scattering at the semiconductor/oxide interface dominates and the mobility decreases\cite{Sabbagh_industrial}. The uniform mobility measured across devices at high density points to a semiconductor/oxide interface with uniform properties across the wafer. Peak mobility and percolation density characterize disorder in the system at high and low density, respectively\cite{Sabbagh_industrial,Lodari_low_2021}.
Sample A (dark blue) and B (blue) show very high peak mobility ($\mu_{max} = 17.6$ and $17.4 \times 10^3$~cm$^2$/Vs, respectively) at low density ($n_H = 5.75$ and $4.96 \times 10^{11}$~cm$^{-2}$, respectively). The inset in Fig.~\ref{fig:1}(b) shows a box plot of the peak mobility across the devices, with an average peak mobility $\mu_{max} = (12.9 \pm 3.4) \times 10^{3}$~cm$^{2}$/Vs. The percolation density $n_p$ is extracted from a percolation fit\cite{Tracy2009} of the density-dependent conductivity $\sigma_{xx} \sim (n_H - n_p)^{p}$ [Fig.~\ref{fig:1}(c)], Device A), with $p=1.31$ fixed for a 2D systems\cite{Tracy2009}\footnote{Since the percolation theory is only valid close to the turn-on threshold, a fitting density range as small as $1.3 \times 10^{11}$~cm$^{-2}$ is used}. The inset in Fig.~\ref{fig:1}(c) shows a box plot of the obtained $n_p$ for all devices. We obtain a very low minimum $n_p$ of $(3.5 \pm 0.4) \times 10^{10}$~cm$^{-2}$ with an average value of $(8.7 \pm 4.0) \times 10^{10}$~cm$^{-2}$. Overall, the maximum mobility in these Hall bars matches the highest values reported for Si-MOS devices with sub 10~nm oxide thickness from ref.~\cite{camenzind2021highmob}. Most importantly, we set the benchmark for percolation density, which is the significant metric for disorder since quantum dot qubits operate in the low-density regime\cite{camenzind2021highmob,Kim_lowPercolation2017}.

We now proceed to evaluate valley splitting from the quantum interference properties of magnetotransport. We focus on three devices (A, B, and C), with various degrees of disorder characterized by $\mu_{max}$ in the range of $12.5$~to~$17.6 \times 10^3$~cm$^{2}$/Vs and $n_p$ in the range of $6.5$~to~$1.1 \times 10^{10}$~cm$^{-2}$. Figure~\ref{fig:2}(a) shows a typical magneto-resistivity curve from device B measured at $65$~mK and at a density near peak mobility. The longitudinal resistivity $\rho_{xx}$ shows Shubnikov-de Haas oscillations (SdH) and Zeeman splitting at magnetic field $B> 0.6$ and $3.5$~T, respectively. At higher magnetic field the oscillation minimum goes to zero, confirming high-quality quantum transport\footnote{A single particle relaxation time $\tau_q = (0.65 \pm 0.02)$~ps is extracted from the SdH oscillations envelope, corresponding to a Landau levels broadening of $\Gamma = 508~\mu$eV, that is, qualitatively, a lower boundary to the VS that can be resolved in magnetotransport.}. Figure~\ref{fig:2}(b) shows the SdH oscillations amplitude $\Delta\rho_{xx}$ measured at $T=1.7$~K as a function of $1/B$ and at increasing density. At lower densities (dark blue curve), the oscillations are periodic in $1/B$ and their amplitude increases following an exponential envelope, a clear indication of conduction through a single-channel. As the 2DEG density increases (blue to orange curves), the oscillation frequency increases and a beating pattern is developed at a relatively low field ($\sim 2$~T). The beating pattern features a full modulation of the oscillations amplitude at $1.7$~K and reveals nodes with a position shifting toward higher magnetic fields $B$ as $V_g$ increases. This interference pattern is a signature of two parallel channels with similar high mobility contributing to transport. We attribute the origin of these channels to the two low-lying conducting band valleys in Si, since the valley splitting energy should dominate at low fields over cyclotron and Zeeman energy\cite{Pudalov2001VS,CHEREMISIN_2004_VS}.

To quantify the population of the two valleys, we show in Fig.~\ref{fig:2}(c) the normalized fast Fourier transform (FFT) spectra amplitude of the oscillations as a function of $V_g$ and oscillations frequency $f$. For $V_g \leq 4.6$~V we observe a single peak in the FFT spectra, pointing to a similar population of the two valleys ($n_1 \sim n_2$) within the experimental resolution of the FFT. Because valley splitting grows with the valley population difference $\Delta n = n_1-n_2$\cite{ando_electronic_1982,Isihara_1986_fittingmodel,Pudalov2001VS,Takashina_VS_2006}, 

\begin{equation}
\Delta E_{VS} = 2\epsilon_F\Delta n/n
\label{valleysplittingequation}
\end{equation}
where $n$ is the total density, $\epsilon_F=\pi\hbar^2n/m^*$ is the Fermi energy for spin-degenerate states and $m^*$ is the effective mass in silicon, this is the regime characterized by a valley spitting smaller than the disorder-induced Landau level broadening $\Delta E_{VS} \leq \Gamma$ implying that beatings are not resolved in the SdH oscillations. Effectively, we measure transport through a single channel whose total density, and hence peak frequency, increases linearly with $V_g$. For $V_g\geq 4.6$~V we start to observe two distinct peaks at frequencies $f_1$ and $f_2$ because the increasing valley splitting overcomes the Landau level broadening.
Correspondingly, two high mobility channels emerge in transport and beatings appear in the SdH oscillations [Fig.~\ref{fig:2}(b)]. We exclude intersubband resonant scattering\cite{leadley_intersubband_1992,coleridge_inter-subband_1990} and treat these two channels independently since $f_1 \sim f_2$ in our measurements and we don't observe features associated to periodicity $f_1-f_2$.
The frequency separation between peaks $|(f_1-f_2)|$, and hence $|(\Delta n)|$, grows with $V_g$, signaling an increasing valley splitting with electric field, in agreement with theoretical expectations\cite{Fiesen2007VS_theory,Pudalov2001VS,CHEREMISIN_2004_VS}. Figure~\ref{fig:2}(d) shows the $V_g$-dependent carrier density in the two valleys $n_{1}$ and $n_{2}$ determined by the quantum Hall density vs. peak frequency relationship $n_{1,2} = g_z g_v f_{1,2} (e/h)$, where $e$ and $h$ are the electron charge and the Planck's constant, $g_z=2$ and $g_v=1$ are Zeeman and valley degeneracy\footnote{When the valley splitting is not resolved ($V_g < 4.6$~V) we assume equal population of the two valleys}. The total density determined by the FFT analysis of the SdH oscillations $ n_{1} + n_{2}$ (black) matches the Hall density $n_H$ (red) obtained at low fields, confirming the validity of the two-band model for transport.

To extract valley splitting we use the following procedure. We identify the SdH oscillations showing clear beatings and fit the curves building upon the models reported in refs.~\cite{Isihara_1986_fittingmodel,Pudalov2001VS} that describe the quantum oscillations at low/intermediate magnetic fields with Lifshitz-Kosevich formulae\cite{LIFSHITS_KOSEVICH_formulae}. The normalized oscillatory part of the magnetoresistance $\Delta \rho_{xx}/\rho_{0}$ is modelled as:

\begin{equation}
\frac{\Delta\rho_{xx}}{\rho_0} = AC(\tau_q) \left[\text{cos}(\frac{\beta}{B}n_1-\pi) + \text{cos}(\frac{\beta}{B}n_2-\pi)\right], 
\label{fit}
\end{equation}
where $A$ is an amplitude prefactor that includes the spin degeneracy, C($\tau_q$) a term depending on the single particle relaxation time $\tau_q$, as detailed below, and $\beta=\pi h/ e$ a constant term. We use three fitting parameter. The first two fitting parameters are $A$ and $\tau_q$, that enters Eq.~\ref{fit} via the temperature dependent term 
\begin{equation}
C(\tau_q) = e^{-\frac{\pi}{\omega_c \tau_q}} \cdot \frac{2\pi^2k_BT/\hbar\omega_c}{sinh(2\pi^2k_BT/\hbar\omega_c)} cos(\frac{\pi g^{*}m^{*}}{2m_0}), 
\label{C(tauq)}
\end{equation}
where we assume a spin susceptibility $g^{*}m^{*}/m_0=0.38$ in silicon and $\tau_q$ equal for the two valleys to minimize the number of fitting parameters, $\omega_c$ is the cyclotron frequency, $k_B$ and $\hbar$ are the Boltzmann and Planck's constants. The third fitting parameter is the valley population difference $\Delta n$ that enters Eq.~\ref{fit} via the two valleys population $n_1 = (n + \Delta n)/2$ and $n_2 = (n -\Delta n)/2$, where $n$ is the total density of the two valleys from Fig.~\ref{fig:2}(d). Whilst $A$ and $C$ in Eq.~\ref{fit} capture the overall shape of the curve, $\Delta n$ influences the periodicity of the SdH oscillations and the interference patterns and is the key parameter to determine valley splitting via Eq.~\ref{valleysplittingequation}\footnote{As initial guess of the fitting procedure we use the density difference $\Delta n = |n_1 - n_2|$ values obtained via FFT, $A = 1$, and $\tau_q = 0.5$~ps for the valley population difference, the amplitude, and the single particle relaxation time, respectively. Since $A$ and $\tau_q$ are not relevant for the valley splitting determination, they are not further discussed in the following of this work. Nonetheless, values of $\tau_q$ compatible with those extracted from the Dingle plot close to peak mobility are found in the range 0.42--0.16~ps for increasing densities.}.

Figure~\ref{fig:3}(a) shows, as an example of the fitting procedure, the experimental data $\Delta\rho_{xx}/\rho_0$ (black) from sample B as a function of $1/B$, measured at high density ($n = 6.75 \times 10^{12}$~cm$^{-2}$) and at $1.7$~K. Beating nodes are observed at $\sim0.23$~T$^{-1}$ ($\sim4.45$~T) and at $\sim0.38$~T$^{-1}$ ($\sim2.63$~T) and the fitted curve (red) matches well the experimental data. The fitted valley population difference $\Delta n = 3.2 \times 10^{11}$~cm$^{-2}$ corresponds to a valley splitting energy $\Delta E_{VS} = 8.2$~meV. To improve on the temperature broadening of the oscillations, and better resolve the beatings, the sample is also cooled down to $65$~mK. Fig.~\ref{fig:3}(b) shows the experimental data (black) and the corresponding fitting (blue) at a lower density ($n = 3.93 \times 10^{12}$~cm$^{-2}$). The beatings nodes are better resolved and from the fitting parameter $\Delta n = 1.64 \times 10^{11}$~cm$^{-2}$ we obtain $\Delta E_{VS} = 4.1$~meV, respectively\footnote{We do not see an increase of the amplitude modulation of beatings in the Shubnikov de-Haas oscillations with increasing temperature, confirming the absence of intersubband resonant scattering.}. 
Figure~\ref{fig:3}(c) summarizes the results of our fitting procedure and shows the valley splitting for Si-MOS devices A (circles), B (squares), and C (asterisks) estimated at $T = 1.7$~K (red) and $65$~mK (blue) as a function of density in the range of $3.7$ to $6.8 \times 10^{12}$~cm$^{2}$. We compare our results with the experimental results (magenta diamonds, from ref.~\cite{Paquelet_VS}) and effective mass calculations (black line, from ref.~\cite{Fiesen2007VS_theory}) for valleys splitting in 2DEGs obtained in Si/SiGe heterostructures. Note that in the Si/SiGe heterostructures in ref.~\cite{Paquelet_VS} valley splitting was estimated by activation measurements in the quantum Hall regime. In Si-MOS, we observe large valley splitting energies that increase in the range of $3.7$ to $8.2$~meV near-linearly with density, regardless of the device location on the wafer and temperature. This in agreement with the observation of a uniform mobility at high density across devices pointing to a uniform semiconductor-dielectric interface across the wafer.

Crucially, we see that the valley splitting density-dependence in Si-MOS extends to the high density regime the same trend that was observed in Si/SiGe at low density\cite{Paquelet_VS}. This trend is compatible\footnote{A linear fit of all the $\Delta E_{VS} $ data for Si-MOS results in a slope for the valley splitting density dependence of $(1.15 \pm 0.02)$~meV$\cdot 10^{-12}$~cm$^{2}$.
The linear fitting intercept has been fixed to 0.} with the predicted density-dependent valley splitting calculated for a disorder-free Si/SiGe quantum well top-interface\cite{Fiesen2007VS_theory}. From this observation we arrive at the following learning: the electron density, and hence the vertical electric field is the key parameter determining the measured valley splitting of 2DEGs in silicon in these wafers, regardless of the interface providing quantum confinement (Si/oxide in Si-MOS or Si/SiGe). We speculate that this apparent universal dependence of valley splitting upon density, and hence electric field, emerges in Hall bar measurements for the following two reasons. Firstly, in a 2DEG the electric field is accurately determined as it connects directly to the measured density. Secondly, macroscopic Hall bar measurements average out the locally varying atomic-scale features at the confining interface that influence valley splitting variations in quantum dots.

In conclusion, we measured the density-dependent valley splitting in Si-MOS Hall-bar transistors made by advanced semiconductor manufacturing. Low-disorder in these Hall-bars allow to estimate valley splitting by analyzing the beating patterns arising from the two energy split valleys in magnetotransport. Comparing the data with previous theory and experimental work for 2DEGs in Si, we highlight the critical role of vertical electric field in determining valley splitting.\\

Data sets supporting the findings of this study are
available at https://doi.org/10.4121/17136608.v1

\bibliography{main.bib}

\end{document}